\documentclass[prl,twocolumn,
superscriptaddress,
longbibliography,
aps
]{revtex4-1}
\usepackage{graphicx}
\usepackage{amsmath}
\usepackage{amsfonts}
\usepackage{amssymb}
\usepackage[colorlinks=true,linkcolor=blue,urlcolor=blue,citecolor=blue]{hyperref}
\usepackage{color}

\newcommand{\beq}{\begin{equation}}
	\newcommand{\eeq}{\end{equation}}

\usepackage{dsfont}
\usepackage{placeins} %%% 02/12/14
\usepackage{bm}           % % % 05/12/14

\usepackage{times}

\usepackage{scalerel}
\usepackage{tikz}
\usetikzlibrary{svg.path}
\definecolor{orcidlogocol}{HTML}{A6CE39}
\tikzset{
	orcidlogo/.pic={
		\fill[orcidlogocol] svg{M256,128c0,70.7-57.3,128-128,128C57.3,256,0,198.7,0,128C0,57.3,57.3,0,128,0C198.7,0,256,57.3,256,128z};
		\fill[white] svg{M86.3,186.2H70.9V79.1h15.4v48.4V186.2z}
		svg{M108.9,79.1h41.6c39.6,0,57,28.3,57,53.6c0,27.5-21.5,53.6-56.8,53.6h-41.8V79.1z M124.3,172.4h24.5c34.9,0,42.9-26.5,42.9-39.7c0-21.5-13.7-39.7-43.7-39.7h-23.7V172.4z}
		svg{M88.7,56.8c0,5.5-4.5,10.1-10.1,10.1c-5.6,0-10.1-4.6-10.1-10.1c0-5.6,4.5-10.1,10.1-10.1C84.2,46.7,88.7,51.3,88.7,56.8z};
	}
}
\newcommand\orcid[1]{\href{https://orcid.org/#1}{\mbox{\scalerel*{
				\begin{tikzpicture}[yscale=-1,transform shape]
					\pic{orcidlogo};
				\end{tikzpicture}
			}{R}}}}

\begin{document}
	\title{Lindblad master equations for quantum systems coupled to dissipative bosonic modes}
	\author{Simon~B.~J\"ager\,\orcid{0000-0002-2585-5246}}
	\affiliation{Physics Department and Research Center OPTIMAS, Technische Universit\"at Kaiserslautern, D-67663, Kaiserslautern, Germany}
	\affiliation{JILA and Department of Physics, University of Colorado, Boulder, Colorado 80309-0440, USA.}
	\author{Tom~Schmit\,\orcid{0000-0001-9904-5417}}
	\affiliation{Theoretical Physics, Department of Physics, Saarland University, 66123 Saarbrücken, Germany}
	\author{Giovanna Morigi\,\orcid{0000-0002-1946-3684}}
	\affiliation{Theoretical Physics, Department of Physics, Saarland University, 66123 Saarbrücken, Germany}
	\author{Murray~J.~Holland\,\orcid{0000-0002-3778-1352}}
	\affiliation{JILA and Department of Physics, University of Colorado, Boulder, Colorado 80309-0440, USA.}
	\author{Ralf~Betzholz\,\orcid{0000-0003-2570-7267}}
	\email{ralf\_betzholz@hust.edu.cn}
	\affiliation{School of Physics, International Joint Laboratory on Quantum Sensing and Quantum Metrology, Institute for Quantum Science and Engineering, Huazhong University of Science and Technology, Wuhan 430074, China}
	
	\begin{abstract}
		We present a general approach to derive Lindblad master equations for a subsystem whose dynamics is coupled to dissipative bosonic modes. The derivation relies on a Schrieffer-Wolff transformation which allows to eliminate the bosonic degrees of freedom after self-consistently determining their state as a function of the coupled quantum system. We apply this formalism to the dissipative Dicke model and derive a Lindblad master equation for the atomic spins, which includes the coherent and dissipative interactions mediated by the bosonic mode. This master equation accurately predicts the Dicke phase transition and gives the correct steady state. In addition, we compare the dynamics using exact diagonalization and numerical integration of the master equation with the predictions of semiclassical trajectories.  We finally test the performance of our formalism by studying the relaxation of a NOON state  and show that the dynamics captures quantum metastability beyond the mean-field approximation.
	\end{abstract}

	\maketitle
	
\begin{figure}[tb]
		\center
		\includegraphics[width=1\linewidth]{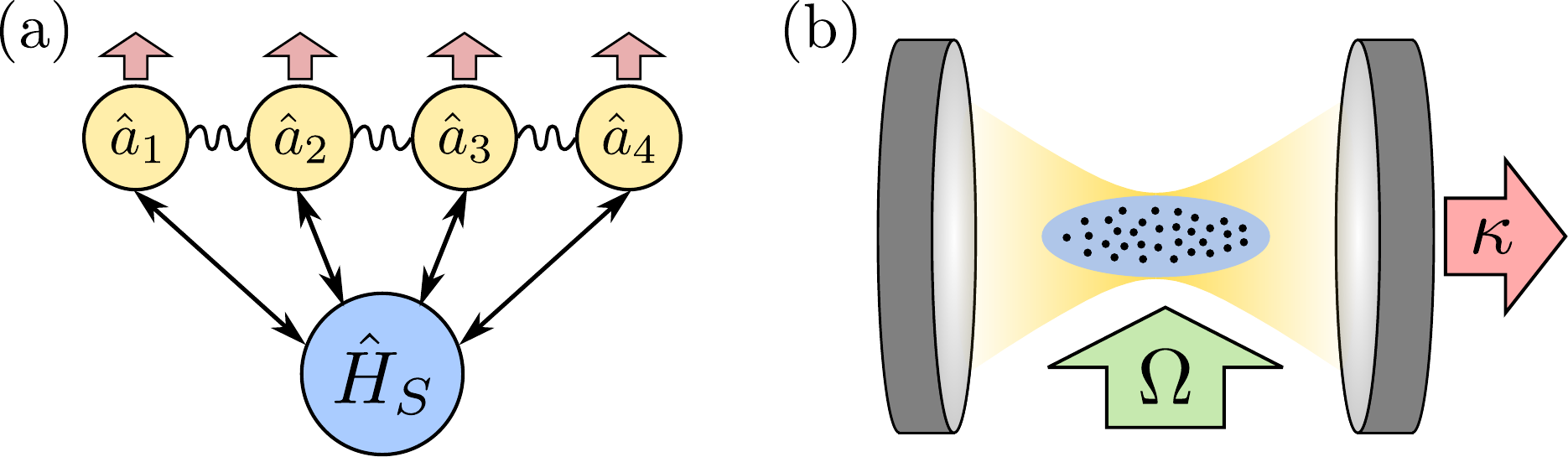}
		\caption{(a) The general model includes coupled dissipative bosonic modes $\hat{a}_k$ interacting with a quantum system described by $\hat{H}_S$. (b) Example: a dissipative optical cavity mode couples to a cloud of driven atoms.\label{Fig:1}}
	\end{figure}

	\textit{Introduction.---}
	The description of open many-body quantum systems dynamics is a formidable challenge for modern experimental and theoretical physics. A typical out-of-equilibrium scenario is theoretically described by a quantum system (QS) which interacts with an environment composed of bosonic modes (BM)~\cite{Leggett:1987} [see Fig.~\ref{Fig:1}(a)]. This is the common setup of quantum electrodynamics,  where the BM are the electromagnetic field~\cite{Gross:1982,Heitler:1984}. Furthermore, it is also at the basis of prominent implementations of quantum simulators because it allows one to tailor the interactions between the constituents of the QS~\cite{Barreiro:2011,Schneider:2012,Monroe:2019,Periwal:2021}. Examples include quantum gases in optical cavities~\cite{Baumann:2010,Gopalakrishnan:2011,Ritsch:2013, Mivehvar:2021,Periwal:2021}, optomechanical arrays~\cite{Heinrich:2011}, phonon-mediated interactions of trapped ions~\cite{Barreiro:2011,Monroe:2019,Schneider:2012,Ge:2019}, polaritons or nitrogen-vancancy centers in diamond coupled to microcavities or mechanical elements~\cite{Rodriguez:2016, Angerer:2018,Cao2017}, and photonic crystals~\cite{Yu:2019}. 
	
	A powerful theoretical tool to analyze open many-body QS is provided by the Keldysh approach~\cite{DallaTorre:2013,Sieberer:2016}, which uses the toolbox of modern quantum field theory to obtain numerical and analytical results. These methods are very successful in predicting the asymptotic behavior of the open QS. The description of dynamics and metastability is instead accessed by full simulations or so-called effective master equations. The latter dispose of a large part of the Hilbert space by eliminating the BM's degrees of freedom~\cite{Schuetz:2013,Damanet:2019,Palacino:2021,Bezvershenko:2021} and include the interactions, noise, and dissipation mediated by them. The derivation of effective master equations are an active research field~\cite{Englert:2002} with a variety of emphases, such as high-precision metrology~\cite{Buchheit:2016,Konovalov:2020}, exact solutions~\cite{Betzholz:2014,Torres:2019,Betzholz:2020}, multi-mode configurations~\cite{Nimmrichter:2010,Torggler:2017,Keller:2017,Keller:2018}, and dynamics of coherent many-body systems~\cite{Damanet:2019,Palacino:2021,Bezvershenko:2021}.
	
	Recently, in the field of cavity many-body quantum electrodynamics, effective Redfield master equations have been derived~\cite{Damanet:2019,Palacino:2021}. While describing the correct low-frequency behavior they are not necessarily completely positive. Attempts to make these master equations positive, e.g., by bringing them into Lindblad form, resulted in predictions of the incorrect asymptotic behavior. Other approaches use effective descriptions by adding fluctuations around a mean-field treatment of the cavity field~\cite{Bezvershenko:2021}. Here, the problem of positiveness was resolved by making a thermalization assumption for the QS which is questionable regarding the possibility of non-thermal metastable states on intermediate timescales~\cite{Schuetz:2016}. These attempts highlight the need to identify general effective descriptions that preserve the positivity. With such effective and positive descriptions one could for instance determine the spectrum of the open system or simulate the master equation using a quantum state diffusion model~\cite{Gisin:1992}. This can then be used to analyze the critical properties of driven-dissipative QS~\cite{Diehl:2010,Minganti:2018,Soriente:2021}, to study prethermalization and metastability~\cite{Schuetz:2016,Macieszczak:2016}, and to shed light on several aspects of the dynamics that cannot be accessed easily otherwise, including measurement-induced phase transitions~\cite{Mueller:2022,Block:2022,Minato:2022}. 
	
	In this Letter, we identify a general procedure which allows derivation of effective master equations for an arbitrary QS that is coupled to dissipative BM. We use a specific type of Schrieffer-Wolff transformation~\cite{Bravyi2011} to reduce the coupling between the QS and the BM such that we can eliminate the latter. This transformation is a displacement that depends, in general, on the eigenstates and eigenenergies of the decoupled QS to be analyzed. The resulting master equation has the Lindblad form, which guarantees positivity if the jump operators are bounded, and the specific procedure allows us to systematically include the effects of retardation between the QS and BM. As an example, we derive an effective, atom-only Lindblad master equation for the dissipative Dicke model and benchmark our results by comparing the spectrum and dynamics with the one of the composite system. 
	
	\textit{Derivation of the effective master equation.---}
	We start by considering a set of BM, described by the annihilation (creation) operators $\hat{a}_k$  ($\hat{a}_k^{\dag}$), with eigenenergies $\omega_k$, that thermalize rapidly at the finite rate $\kappa_k$ with an external bath with inverse temperature $\beta$. The dynamics for the density matrix $\hat{\rho}$ is described by $    {\mathcal{L}_{\mathrm{d}}\hat{\rho}=\sum_k\{\kappa_k(n_k+1)\mathcal{D}[\hat{a}_k]\hat{\rho}+\kappa_kn_k\mathcal{D}[\hat{a}_k^{\dag}]\hat{\rho}\}},$
	where we introduced $\mathcal{D}[\hat{O}]\hat{\rho}=2\hat{O}\hat{\rho}\hat{O}^{\dag}-\hat{O}^{\dag}\hat{O}\hat{\rho}-\hat{\rho}\hat{O}^{\dag}\hat{O}$. In this Letter, we consider the case where $n_k=[\exp(\beta\omega_k)-1]^{-1}\approx0$, which is usually valid if the $\omega_k$ are optical frequencies. Throughout this work we use $\hbar=1$. On a timescale that is longer than $1/\kappa_k$, the BM couple coherently to a QS described by the Hamiltonian 
	\begin{equation}
		\hat{H}=\hat{H}_S+\sum_k\left(\sum_{k'}\hat{a}^{\dag}_{k'}\hat{\Omega}_S^{k',k}\hat{a}_k+\hat{a}^{\dag}_k\hat{S}_k+\hat{S}_k^{\dag}\hat{a}_k	\right).\label{H}
	\end{equation}
	The Hamiltonian in absence of the BM is denoted by $\hat{H}_S$. The term proportional to $\hat{\Omega}_S^{k',k}=(\hat{\Omega}_S^{k,k'})^{\dag}$ denotes the frequencies and mode-mode coupling that may depend on the QS's degrees of freedom. The last term, $\hat{a}^{\dag}_k\hat{S}_k+\hat{S}_k^{\dag}\hat{a}_k$, represents the driving of the BM, which may likewise include QS operators, described by $\hat{S}_k$ and $\hat{S}^{\dag}_k$. The dynamics of the density matrix $\hat{\rho}$ for the combination of both the BM and QS is then described by the master equation
	\begin{equation}
		\frac{\partial \hat{\rho}}{\partial t}=\mathcal{L}\hat{\rho}:=-i[\hat{H},\hat{\rho}]+\mathcal{L}_{\mathrm{d}}\hat{\rho}.\label{Mastereq0}
	\end{equation}
	
	We want to eliminate the BM degrees of freedom and derive an effective master equation that describes the dynamics of the QS.  The steps for the derivation are as follows: (i) We first derive the master equation for $\tilde{\rho}=\hat{D}^{\dag}\hat{\rho}\hat{D}$  where $\hat{D}=\exp[\sum_k(\hat{a}^{\dag}_k\hat{\alpha}_k-\hat{\alpha}_k^{\dag}\hat{a}_k)]$ is the displacement operator that weakly correlates the BM $\hat{a}_k$ to the QS by establishing an effective-field operator $\hat{\alpha}_k$. We assume that $\|\hat{\alpha}_k\|\sim\epsilon\ll1$, therefore, we can apply a perturbation theory where we discard all terms that are of third order in $\epsilon$ or higher. (ii) In the displaced picture, we project the BM onto the thermal state that is in our parameter regime essentially the vacuum state $|\text{vac}\rangle$ and define $\hat{\rho}_{\text{sys}}=\langle \text{vac}|\tilde{\rho}|\text{vac}\rangle$. We systematically include the coupling of $\hat{\rho}_{\text{sys}}$ to higher Fock states in the displaced BM and optimize the operators $\hat{\alpha}_k$, that we denote as effective fields, such that $\hat{\rho}_{\text{sys}}$ is decoupled up to third order in $\epsilon$. This decoupling procedure is reminiscent to a Schrieffer-Wolf transformation. In the Supplemental Material (SM)~\cite{SM} we show that these steps result in solving
	\begin{equation}
		\label{alpha_k}
		\frac{\partial \hat{\alpha}_k}{\partial t}	=-i[\hat{H}_S,\hat{\alpha}_k]-i\sum_{k'}\hat{\Omega}_S^{k,k'}\hat{\alpha}_{k'}-i\hat{S}_k-\kappa_k\hat{\alpha}_k.
	\end{equation}
	With the solution $\hat{\alpha}_k$ of the above equation, we obtain a master equation for the density matrix $\hat{\rho}_{\text{sys}}$ that reads
	\begin{equation}
		\frac{\partial\hat{\rho}_{\text{sys}}}{\partial t}=\mathcal{L}_{\text{eff}}\hat{\rho}_{\text{sys}}:=-i[\hat{H}_{\text{eff}},\hat{\rho}_{\text{sys}}]+\sum_k\kappa_k\mathcal{D}[\hat{\alpha}_k]\hat{\rho}_{\text{sys}}\label{effMaster}
	\end{equation}
	and the effective Hamiltonian \begin{align}
		\hat{H}_{\text{eff}}=\hat{H}_S+\frac{1}{2}\sum_{k}(\hat{\alpha}^{\dag}_k\hat{S}_k+\hat{S}_k^{\dag}\hat{\alpha}_k).  
	\end{align}
	
	This master equation is the main result of this Letter that we now discuss in greater depth. We first observe that this effective master equation is of the Lindblad form, thereby preserving the positivity if the $\hat{\alpha}_k$ are bounded operators. Although this derivation was performed for the multimode case, below we mostly focus on the single-mode case~\footnote{For the single-mode case we drop the index $k$ for conciseness}. The terms proportional to $\kappa$ and  $\hat\Omega_S$ in Eq.~\eqref{alpha_k} describe the relaxation of the BM to the thermal state in absence of $\hat{S}$. During this relaxation, the QS is evolving according to $\hat{H}_S$ such that the BM sees a retardation effect determined by $[\hat{H}_S,\hat{\alpha}]$. This term is a principal finding because it shows that the BM carries information about the evolution of the QS which is here determined by $\hat{H}_S$. In fact, solving Eq.~\eqref{alpha_k} for the steady state, assuming that $[\hat{H}_S,\hat{\alpha}]$ can be ignored, results in the adiabatic elimination~\cite{Larson:2008,Larson:2008:2,Habibian:2013} given by $\hat{\alpha}=-i\hat{S}/(i\hat\Omega_S+\kappa)$. This includes quantum fluctuations of the field due to $\kappa$, visible by the proportional incoherent part in  Eq.~\eqref{effMaster}. With this, it also correctly recovers the dispersive limit, $\|\hat\Omega_S\|\gg\kappa$, where the dynamics of the QS, described solely by $\hat{H}_{\mathrm{eff}}$, evolves coherently. Using Eqs.~\eqref{alpha_k} and \eqref{effMaster}, we can now systematically take retardation and noise effects into account by treating $[\hat{H}_S,\hat{\alpha}]$ and $\kappa$ either in arbitrary order, or as a perturbation. We remark here, that first-order perturbative expansions in retardation effects have been studied in semiclassical descriptions, giving rise to collective cavity cooling and dissipation-assisted prethermalization~\cite{Schuetz:2013,Schuetz:2016,Xu:2016,Jaeger:2017,Keller:2017,Keller:2018}. However, the effective master equation~\eqref{effMaster} is a full quantum description and therefore complementary to the results of Refs.~\cite{Damanet:2019,Palacino:2021,Bezvershenko:2021} that derive effective quantum descriptions. Similar to Ref.~\cite{Bezvershenko:2021} we use a displacement operation to eliminate the BM, now, however, this ``displacement'' is not based on an underlying mean-field assumption, but instead ``$\hat{\alpha}$'' is an operator that intrinsically includes fluctuations. Our approach requires thermalization of the BM, while we do not require thermalization of the QS which allows Eq.~\eqref{effMaster} to describe metastable dynamics. To show the potential of Eqs.~\eqref{alpha_k} and \eqref{effMaster} we will now analyze a particular example, namely the dissipative Dicke model.  
	
	\textit{Application to the dissipative Dicke model.---} The dissipative Dicke model describes the quantum dynamics of a single mode coupled to $N$ two-level atoms. It can be experimentally realized with driven atoms interacting with an optical cavity mode~\cite{Baumann:2010,Dimer:2007} [see Fig.~\ref{Fig:1}(b)]. We therefore denote the QS by the atoms and the BM by the cavity mode.  
	With our definitions in Eq.~\eqref{H} we use $\hat{H}_S=\omega_0 \hat{S}^{z}$, the cavity frequency $\hat{\Omega}_S=\omega_c$, and coupling $\hat{S}=2g\hat{S}^x/\sqrt{N}$. We have introduced the collective spin operators $\hat{S}^a =\sum_{j=1}^N\hat{\sigma}_j^a/2$ with $a\in\{x,y,z\}$ and where $\hat{\sigma}^a_j$ denote the Pauli matrices of the $j$th atom. The dissipative Dicke model exhibits a phase transition in the thermodynamic limit $N\to\infty$ from a normal ($g<g_c$) to a superradiant phase ($g>g_c$)~\cite{Hepp1973a,Hepp1973b,Dimer:2007,DallaTorre:2013}, where the critical value $g_c$ is given by $g_c^2=\omega_0(\omega_c^2+\kappa^2)/(4\omega_c)$. In contrast to the quantum phase transition of the Dicke model~\cite{Hepp1973a,Hepp1973b}, the dissipative Dicke model exhibits different critical exponents and a damping rate at steady state~\cite{Nagy:2011,DallaTorre:2013,Brennecke:2013}. Therefore, it provides an important check as to whether Eq.~\eqref{effMaster} can predict the correct dynamics and steady state.
	
	In Ref.~\cite{Damanet:2019}, it was shown that an atom-only Redfield master equation gives the correct low-frequency behavior of the dissipative Dicke model. In addition, it was demonstrated that this cannot be achieved using an adiabatic elimination or a secularized Linblad master equation. The latter is obtained by dropping the co-rotating and off-resonant $\hat{a}^{\dag}\hat{S}^+$ and $\hat{a}\hat{S}^-$ terms ($\hat{S}^{\pm}=\hat{S}^x\pm i\hat{S}^y$). Based on that analysis it was conjectured that correct, atom-only master equations for the dissipative Dicke model require a non-Lindblad form. We will now show that the Lindblad master equation~\eqref{effMaster} goes beyond the adiabatic and secularization approximation and is a counter example for this conjecture.
	
	We first determine $\hat{\alpha}$ using the commutation relation $	[\hat{S}^a,\hat{S}^b]=i\sum_{c\in\{x,y,z\}}\epsilon_{abc}\hat{S}^c$, where $\epsilon_{abc}$ is the Levi-Civita symbol. The steady state of Eq.~\eqref{alpha_k} is given by
	\begin{equation}
		\hat{\alpha}=\alpha_+\hat{S}^{+}+\alpha_-\hat{S}^{-}\label{alphaDicke},
	\end{equation}
	with $\alpha_{\pm}=-g/[\sqrt{N}(\omega_c\pm\omega_0-i\kappa)]$. As a result of the commutator term $[\hat{H}_S,\hat{\alpha}]$ we find that the effective cavity field $\hat{\alpha}$ has two sidebands shifted by $\pm\omega_0$ from $\omega_c$ that correspond to the excitation or de-excitation of the atoms. If we impose $\omega_0=0$ in Eq.~\eqref{alphaDicke} we recover the adiabatic elimination as in Ref.~\cite{Damanet:2019}. In addition, using Eq.~\eqref{alphaDicke} in Eq.~\eqref{effMaster} we also find co-rotating terms $[\hat{S}^{\pm}]^2$, dropping the latter would result in the secularization approximation of Ref.~\cite{Damanet:2019}. As a first check, we now compare the spectrum of the effective master equation~\eqref{effMaster} with the one of the full master equation~\eqref{Mastereq0} for small atom numbers $N$. To do this, we diagonalize $\mathcal{L}$ and $\mathcal{L}_{\mathrm{eff}}$ using the symmetric atomic states $|m\rangle$, with $\hat{S}^z|m\rangle =m|m\rangle$ for $m=-N/2,-N/2+1,\dots,N/2$, as a basis. 
	
\begin{figure}[tb]
		\center
		\includegraphics[width=1\linewidth]{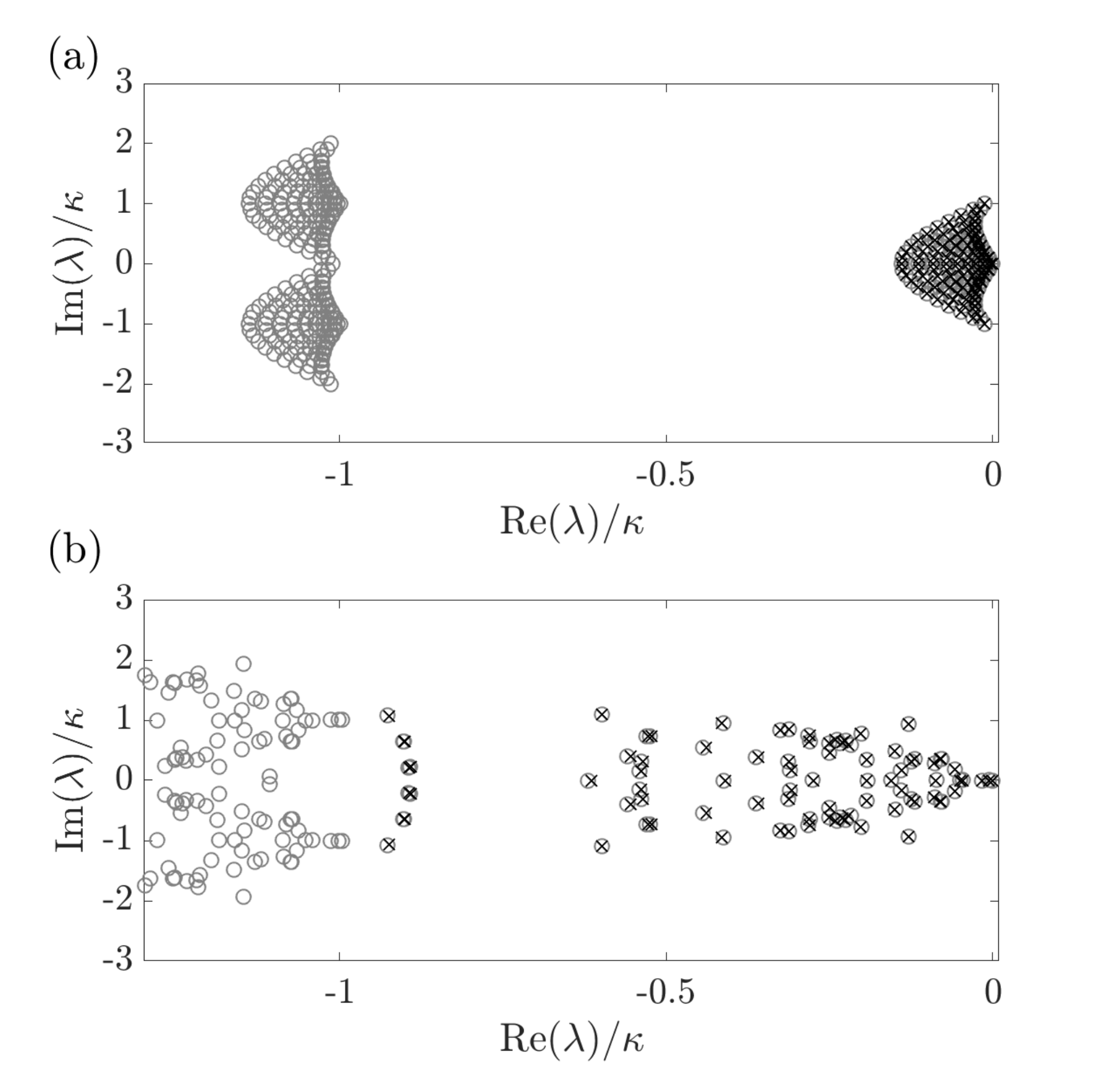}
		\caption{Eigenvalues $\lambda$ in units of $\kappa$ for the master equation~\eqref{Mastereq0} (gray ``o'') and effective master equation~\eqref{effMaster} (black ``x'') for the dissipative Dicke model. The parameters are $N=10$, $\omega_c=\kappa$, $\omega_0=0.1\kappa $, and (a) $g=0.5g_c$, (b) $g=2g_c$.\label{Fig:2}}
	\end{figure}

	In Fig.~\ref{Fig:2}, we show the complex eigenvalues $\lambda$ of $\mathcal{L}$ and $\mathcal{L}_{\text{eff}}$ as gray circles and black crosses, respectively.
	Figure~\ref{Fig:2}(a) represents a parameter choice below threshold, $g<g_c$. In this case, the cavity field is to good approximation in the vacuum state and we find excellent agreement of the full and effective descriptions for the eigenvalues with the largest real parts. This emphasizes that the effective master equation correctly describes long timescales and discards faster timescales with $\text{Re}(\lambda)<-\kappa$, thereby describing the dynamics of the metastable states. Figure~\ref{Fig:2}(b) shows the spectrum for large coupling $g>g_c$, in the superradiant phase. Again, we find very good agreement between the full and effective descriptions, which is remarkable since the gap between the ``correctly'' described modes and $\text{Re}(\lambda)\approx-\kappa$ is much smaller. Nevertheless, the effective master equation still correctly describe the metastable dynamics. This direct comparison suggests that the effective description is valid across the phase transition point of the dissipative Dicke model.
	
\begin{figure}[tb]
		\center
		\includegraphics[width=1\linewidth]{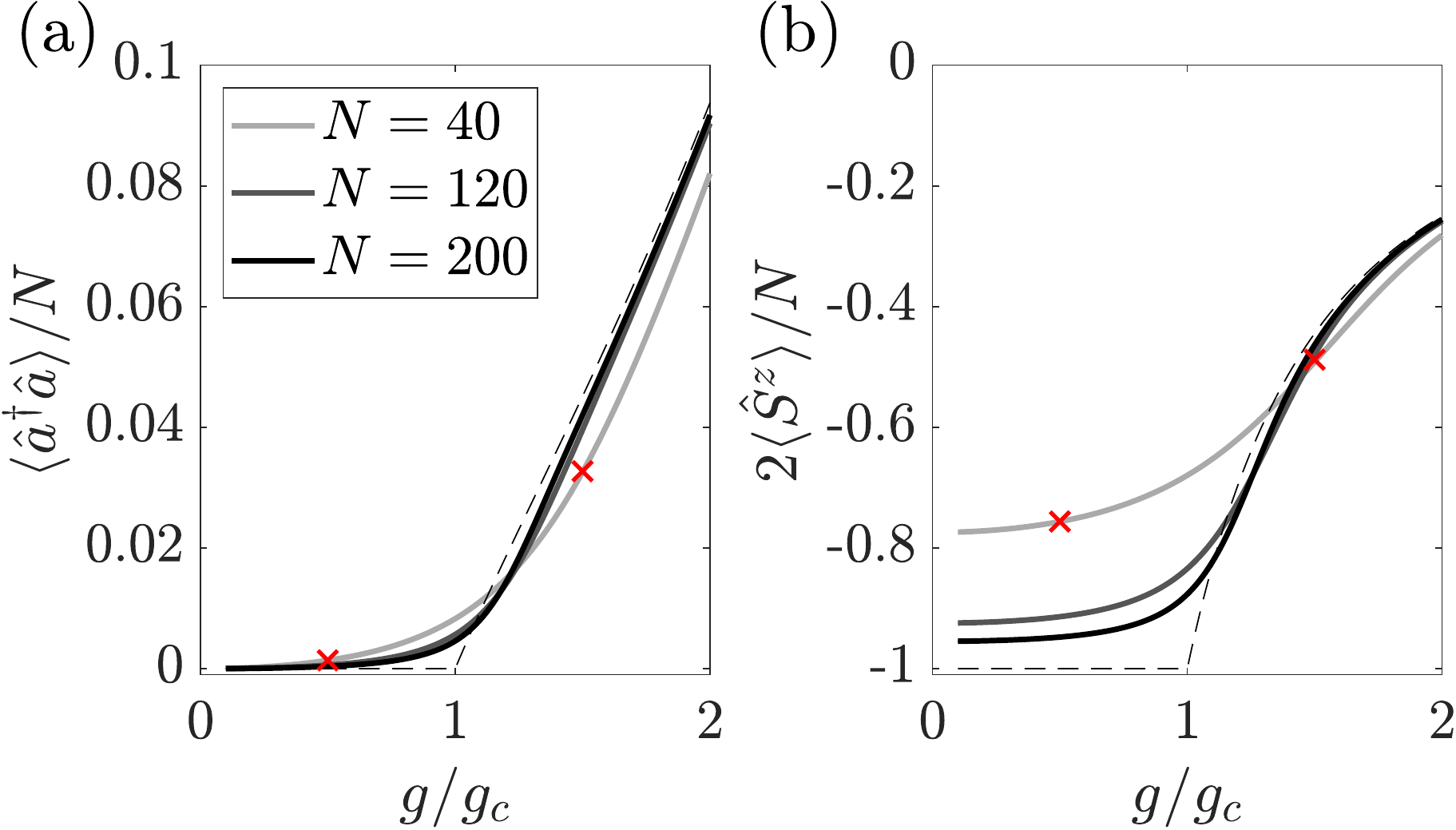}
		\caption{ (a) Photon number $\langle\hat{a}^{\dag}\hat{a}\rangle$ and (b) inversion $\langle\hat{S}^{z}\rangle$ as a function of $g$ in units of the critical coupling strength $g_c$. Dashed lines are the mean-field results for $N\to\infty$ and solid lines are obtained by finding the steady-state of Eq.~\eqref{effMaster} for the dissipative Dicke model with various atom numbers $N$ (see inset). Red crosses are obtained by finding the steady state of the full master equation~\eqref{Mastereq0} for $N=40$. The remaining parameters are $\omega_c=\kappa$, $\omega_0=0.1\kappa $.\label{Fig:3}}
	\end{figure}

	To further support this claim, we now use Eq.~\eqref{effMaster} to make analytical predictions in the limit $\omega_c,\kappa\gg\omega_0$, i.e., the limit when the cavity evolves much faster than the atoms~\cite{Damanet:2019}. For this case, the commutator term $[\hat{H}_S,\hat{\alpha}\hat]$ can be treated perturbatively and the coefficients in Eq.~\eqref{alphaDicke} can be expanded according to $\alpha_{\pm}=-g/[\sqrt{N}(\omega_c-i\kappa)]\pm g\omega_0/[\sqrt{N}(\omega_c-i\kappa)^2]$.
	In the large $N$ limit, we can derive mean-field equations for $S^a=\langle\hat{S}^a\rangle$ with $a\in\{x,y,z\}$ that are reported in the SM~\cite{SM}. We show that the resulting equations are the same as the ones given in Ref.~\cite{Damanet:2019}. Consequently, we find the same threshold and the correct oscillation and damping rates in the thermodynamic limit. In addition, we also find the correct critical exponents of the dissipative Dicke model (see SM~\cite{SM} for details). The steady-state values of $I=\langle \hat{a}^{\dag}\hat{a}\rangle$ and $S^z$ in the thermodynamic limit are given by $I_0=0$ and $S^z_0=-N/2$ for $g<g_c$ and $I_0= Ng^2(1-g_c^4/g^4)/(\omega_c^2+\kappa^2)$ and $S^z_0=-Ng_c^2/(2g^2)$ for $g>g_c$. 
	In Fig.~\ref{Fig:3}(a) and (b), we show these analytical results of $I_0$ and $S^z_0$ as functions of $g$ as black dashed lines. 
	Furthermore, we present the values $I$ and $\langle \hat{S}^z\rangle$ by numerically finding the steady state of Eq.~\eqref{effMaster} and then calculating
	$\langle \hat{a}^{\dag}\hat{a}\rangle=\text{Tr}[\hat{\alpha}^{\dag}\hat{\alpha}\hat{\rho}_\text{sys}]$ and $\langle \hat{S}^{z}\rangle=\text{Tr}[\hat{S}^{z}\hat{\rho}_\text{sys}]$. Since the effective master equation does not include the cavity degrees of freedom, we are able to diagonalize the Lindbladian for much larger atom numbers. As can be seen in Fig.~\ref{Fig:3}(a) and (b), the analytical result and the numerical results are in better agreement for larger atom numbers $N$. For $N=40$, we were also able to find the steady state of the full master equation~\eqref{Mastereq0}, depicted for two values of $g/g_c$ as red crosses. Again, there is excellent agreement with the effective method, indicating that it is also valid for finite atom numbers. Altogether, these results show that the effective master equation predicts the correct steady state and low-frequency oscillations, damping rates, and critical exponents. 
	
\begin{figure}[tb]
		\center
		\includegraphics[width=1\linewidth]{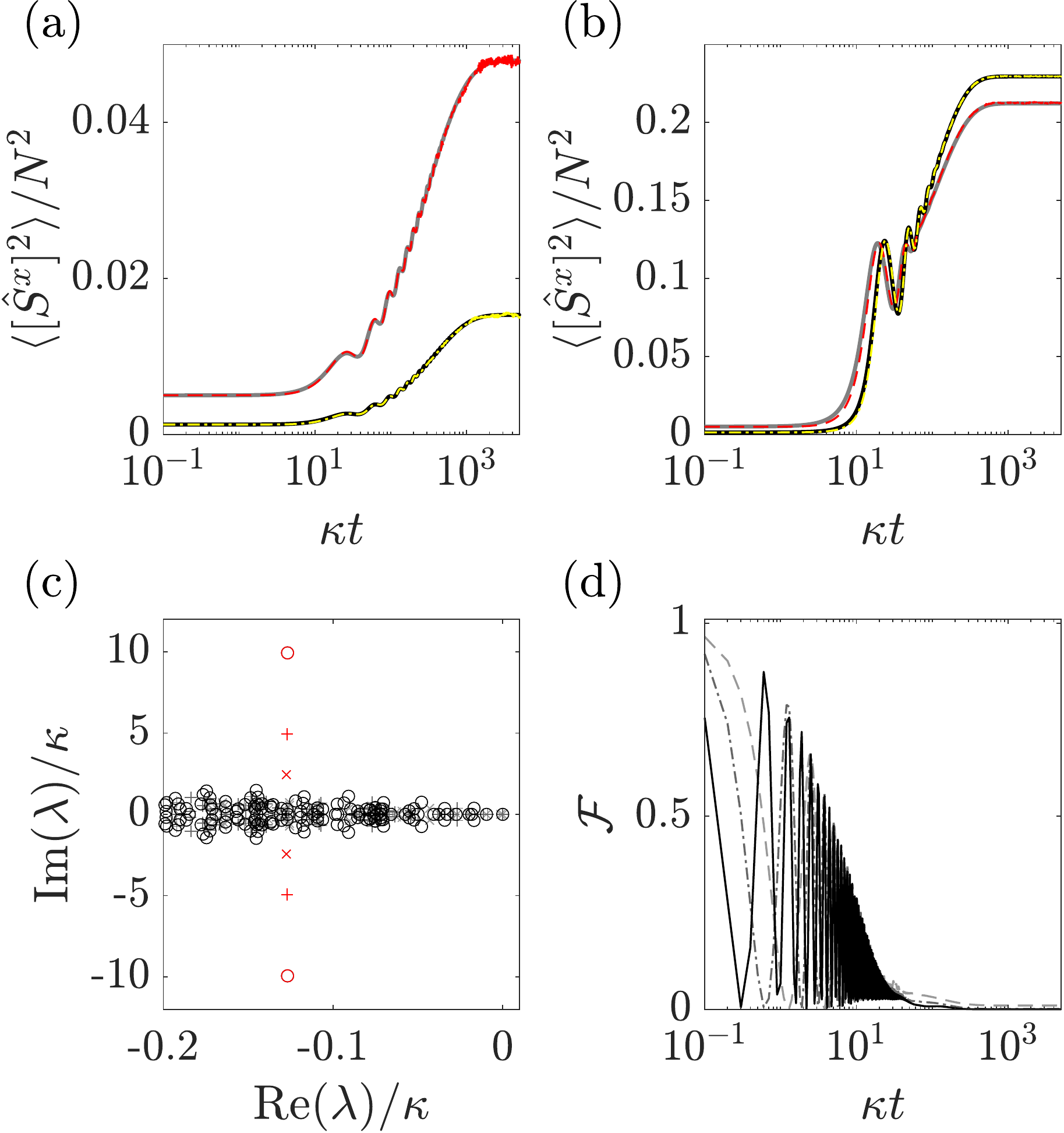}
		\caption{The value of $\langle[\hat{S}^x]^2\rangle$ as function of time in units of $1/\kappa$ for (a) $g=0.5g_c$ and (b) $g=2g_c$. The gray (black) lines are obtained by simulating the effective master equation~\eqref{effMaster} with $N=50$ ($N=200$). The red dashed (yellow dashed-dotted) lines are simulated with the stochastic method reported in the SM~\cite{SM} and averaged over 20000 simulations with $N=50$ ($N=200$). (c) Eigenvalues $\lambda$ in units of $\kappa$ of Eq.~\eqref{effMaster} for $N=25$ (``x''), $N=50$ (``+''), and $N=100$ (``o''). The red symbols mark the eigenvalues discussed in the text. (d) Fidelity $\mathcal{F}=\langle\Psi|\hat{\rho}_{\text{sys}}(t)|\Psi\rangle$ as function of time in units of $1/\kappa$ simulated using Eq.~\eqref{effMaster} initialized with the state $|\Psi\rangle$ discussed in the text for $N=25$ (light gray dashed), $N=50$ (gray dashed-dotted), and $N=100$ (black solid) with $\omega_c=\kappa$, $\omega_0=0.1\kappa $.\label{Fig:4}}
	\end{figure}

	In the last part of this Letter, we focus on the out-of-equilibrium dynamics, i.e., a scenario where the system is initialized ``far'' away from the steady state. The dynamics and relaxation in such a situation require the correct description of high and low frequency excitations and metastable states. Since it is difficult to simulate the full master equation~\eqref{Mastereq0} for large atom numbers, we employ a semiclassical stochastic method to compare with our simulations of the effective master equation~\eqref{effMaster}. The stochastic method simulates the coupled dynamics of the $c$-number equivalents of spin components $S^x,S^y$, and $S^z$ coupled to the noisy real part $x$ and imaginary part $p$ of the field amplitude. Details are reported in the SM~\cite{SM}. In a first benchmark, we initialize the system with all atoms in the ground state, corresponding to a fully polarized state with $\langle\hat{S}^z\rangle=-N/2$, and evolve this state according to Eq.~\eqref{effMaster}. Figure~\ref{Fig:4}(a) and~(b) show the time evolution of $\langle[\hat{S^x}]^2\rangle$ for $g=0.5g_c$ and $g=2g_c$, respectively, for $N=50$ (gray) and $N=200$ (black). 
	These numerical simulations are compared with the stochastic simulations visible as red dashed ($N=50$) and yellow dashed-dotted ($N=200$) lines. We find excellent agreement. Since the stochastic simulations evolve the coupled atom-cavity dynamics on equal footing, we conclude that Eq.~\eqref{effMaster} incorporates the correct retarded interaction between atoms and cavity, and is therefore well suited for the description of out-of-equilibrium dynamics. 
	
	Finally, we want to analyze a scenario with quantum features that cannot be described by the semiclassical stochastic methods~\cite{SM}. To achieve this we first analyze the spectrum of the Eq.~\eqref{effMaster} for $g=2g_c$, which is shown in Fig.~\ref{Fig:4}(c) for $N=25$ (``x''), $N=50$ (``+''), and $N=100$ (``o''). Here, we find a mode with a growing imaginary part for increasing $N$ (marked red). The underlying mode is related to the coherence $\hat{c}=|N/2\rangle\langle-N/2|$ and $\hat{c}^{\dag}$ that oscillate with a frequency $\sim N\omega_0$. Remarkably, the frequency of this mode exceeds the cavity resonance and linewidth while its damping is far less than the cavity linewidth. Therefore, it can be seen as a metastable high-frequency oscillation with a number of coherent periods that diverges for increasing atom number. To find this oscillation dynamically, we initialize the system in the NOON state $|\Psi\rangle=(|N/2\rangle+|-N/2\rangle)/\sqrt{2}$ such that the coherence $\hat{c}$ is present at $t=0$. We then evolve $|\Psi\rangle$ according to the effective master equation~\eqref{effMaster} and calculate the fidelity $\mathcal{F}=\langle\Psi|\hat{\rho}_{\text{sys}}(t)|\Psi\rangle$ visible in Fig.~\ref{Fig:4}(d). We show $\mathcal{F}$ for $N=25$ (light gray dashed), $N=50$ (gray dashed-dotted), and $N=100$ (black solid) and find an oscillation frequency that increases with $N$. At the same time, the relaxation seems to be nearly independent of the particle number, in agreement with Fig.~\ref{Fig:4}(c), where we saw an increasing imaginary part but a nearly constant real part of the red-marked modes. Since the initial state and its dynamics is beyond the mean-field approximation, this further highlights the ability of Eq.~\eqref{effMaster} to describe out-of-equilibrium situations with entangled quantum states. 
	
	\textit{Conclusion.---}We have developed a formalism for the derivation of effective master equations that describe the reduced dynamics of a quantum system coupled to dissipative bosonic modes. This effective master equation is of the Lindblad form, thereby ensuring that the positivity is preserved. Furthermore, our approach includes the retarded interaction between the quantum system and the bosonic modes and therefore goes beyond the often considered adiabatic elimination. We demonstrated this by applying the formalism to the dissipative Dicke model, where it was shown to describe the correct steady state and dynamics for small as well as large atom numbers.
	
	We believe that the method presented here may be extended to the regime where the thermal occupation of the bosonic modes is not negligible. In addition, we expect that it will be possible to generalize this method to include higher coupling strengths, for instance, by modifying the displacement transformation. This might be interesting for systems with a vanishing gap, e.g., atom-cavity systems with a $U(1)$ symmetry~\cite{Palacino:2021} where it was shown that the Redfield master equation is inaccurate. In future, it will be interesting to apply the effective master equation also to multi-mode systems to study many-body cooling, the formation of coherent structures in the presence of dissipation, as well as reservoir and interaction engineering.
	
	%\emph{Acknowledgements} 
	SBJ and MJH acknowledge support from the NSF Q-SEnSE Grant No.~OMA 2016244, NSF PFC Grant No.~1734006, and the DARPA and ARO Grant No. W911NF-16-1-0576. SBJ is thankful for discussions with A. Shankar, J. Reilly, J. Bartolotta, J. Cooper and for support from Research Centers of the Deutsche Forschungsgemeinschaft (DFG): Projects A4 and A5 in SFB/Transregio 185: “OSCAR”. TS and GM acknowledge funding by the Deutsche Forschungsgemeinschaft (DFG, German Research Foundation) Project-ID 429529648 TRR 306 QuCoLiMa (Quantum Cooperativity of Light and Matter) and by the DFG Priority Program No. 1929
	GiRyd.
	
\bibliography{article.bib} 

\newpage 

\setcounter{equation}{0}
\setcounter{figure}{0}
% % %

% % % supp.tex \begin{document}-\end{document}
%\renewcommand*{\citenumfont}[1]{S#1}
\renewcommand*{\bibnumfmt}[1]{[S#1]}
\renewcommand{\thesection}{S~\arabic{section}}
\renewcommand{\thesubsection}{\thesection.\arabic{subsection}}
\makeatletter %% With ams
\def\tagform@#1{\maketag@@@{(S\ignorespaces#1\unskip\@@italiccorr)}}
\makeatother
\makeatletter
\makeatletter \renewcommand{\fnum@figure}
{\figurename~S\thefigure}

\newpage

\onecolumngrid

\newpage
\begin{center}
	\textbf{\large Supplemental Material: Lindblad master equations for quantum systems coupled to dissipative bosonic modes}
\end{center}

\twocolumngrid

	\section{Displaced master equation} \label{App:A}
	In this section, we present some details on the displacement transformation used in the Letter to describe the dynamics of $\tilde{\rho}$. The displacement operator in this transformation, which can be consider as generalization of the Polaron transformation~\cite{Mahan2000}, is defined as
	\begin{equation}
		\hat{D}=\exp\left\{\sum_k[\hat{a}^{\dag}_k\hat{\alpha}_k(t)-\hat{\alpha}_k^{\dag}(t)\hat{a}_k]\right\}\label{D},
	\end{equation}
	where for every bosonic mode (BM) the operator $\hat{\alpha}_k(t)$ acts on the quantum system (QS). We will now apply this displacement operator to the coupled QS-BM master equation, given by $\mathcal{L}$ in the main text. The density matrix in the displaced picture then has the form
	\begin{equation}
		\label{eq:disrho}
		\tilde{\rho}=\hat{D}^{\dag}\hat{\rho}\hat{D}.
	\end{equation}
	We assume that the driving $\hat{S}_k$ is sufficiently small, such that also the displacements $\hat{\alpha}_k$ are sufficiently small. More precisely, we will neglect third-order terms in $\hat{S}_k$ and $\hat{\alpha}_k$, such that $\hat{\alpha}_{k_1}\hat{\alpha}_{k_2}\hat{\alpha}_{k_3}\approx0$ for all possible combinations of $k_1,k_2,k_3$.
	
	The dynamics of the density operator~\eqref{eq:disrho} is determined by the master equation in the displaced picture, which can be written as
	\begin{equation}
		\frac{\partial\tilde{\rho}}{\partial t}=\mathcal{L}_a\tilde{\rho}+\mathcal{L}_b\tilde{\rho},
	\end{equation}
	where the first term, 
	\begin{equation}
		\mathcal{L}_a\tilde{\rho}=\frac{\partial \hat{D}^{\dag}}{\partial t}\hat{D}\tilde{\rho}+\tilde{\rho}\hat{D}^{\dag}\frac{\partial \hat{D}}{\partial t},   
	\end{equation} 
	originates from a possible explicit time dependence of $\hat{\alpha}_k$. The second term,
	\begin{equation}
		\mathcal{L}_b\tilde{\rho}=\hat{D}^{\dag}\frac{\partial \hat{\rho}}{\partial t}\hat{D}=-i[\hat{D}^{\dag}\hat{H}\hat{D},\tilde{\rho}]+\sum_k\kappa_k\tilde{\mathcal{D}}[\hat{a}_k]\tilde{\rho},\label{Displacedmaster}
	\end{equation}
	on the other hand, describes the time evolution in the displaced picture in absence of an explicit time dependence. Here, $\hat{D}^{\dag}\hat{H}\hat{D}$ is the Hamiltonian in the displaced frame, whereas $\tilde{\mathcal D}[\hat{a}_k]\tilde{\rho}=\hat{D}^{\dag}(\mathcal{D}[\hat{a}_k]\hat{\rho})\hat{D}$ represents the displaced dissipator. In the following, we will derive the explicit form of $\mathcal{L}_a$ and $\mathcal{L}_b$.
	
	\begin{widetext}
		\subsection{Calculation of $\mathcal{L}_a$}\label{App:A:1}
		We want to derive the explicit form of $\mathcal{L}_a$. In order to do so, we use $\hat{D}^{\dag}\hat{D}=1$ to establish
		\begin{equation}
			\frac{\partial \hat{D}^{\dag}}{\partial t}\hat{D}=-\hat{D}^{\dag}\frac{\partial \hat{D}}{\partial t}.
		\end{equation}
		As a first step, we want to obtain an expression for the time derivative of the displacement operator. To this end, we write
		\begin{equation}
			\label{eq:dispderiv}
			\frac{\partial\hat{D}}{\partial t}=\frac{\partial}{\partial t}\exp(\hat{r})=\sum_{n=0}^{\infty}\frac{1}{n!}\sum_{m=0}^{n-1}\hat{r}^m\frac{\partial\hat{r}}{\partial t}\hat{r}^{n-m-1},
		\end{equation}
		where we have defined the anti-Hermitian generator $\hat{r}=\sum_{k}(\hat{a}_k^\dag\hat{\alpha}_k-\hat{\alpha}_k^{\dag}\hat{a}_k)$.
		The inner time derivative simply reads
		\begin{equation}
			\frac{\partial \hat{r}}{\partial t}=\sum_{k}\bigg(\hat{a}_k^\dag\frac{\partial \hat{\alpha}_k}{\partial t}-\frac{\partial \hat{\alpha}_k^\dag}{\partial t}\hat{a}_k\bigg)
		\end{equation}
		and can be used to obtain
		\begin{equation}
			\hat{r}^m\frac{\partial\hat{r}}{\partial t}\hat{r}^{n-m-1}=\sum_k\bigg(\hat{a}_k^\dag\frac{\partial\hat{\alpha}_k}{\partial t}\hat{r}^{n-1}+\bigg[\hat{r}^m,\hat{a}^\dag_k\frac{\partial\hat{\alpha}_k}{\partial t}\bigg]\hat{r}^{n-m-1}-\hat{r}^{n-1}\frac{\partial \hat{\alpha}^{\dag}_k}{\partial t}\hat{a}_k-\hat{r}^m\bigg[\frac{\partial \hat{\alpha}_k^{\dag}}{\partial t}\hat{a}_k,\hat{r}^{n-m-1}\bigg]\bigg).
		\end{equation}
		The summation over $m$ in Eq.~\eqref{eq:dispderiv} then results in
		\begin{align}
			\sum_{m=0}^{n-1}\hat{r}^m\frac{\partial\hat{r}}{\partial t}\hat{r}^{n-m-1}=n\sum_{k}\bigg(\hat{a}^\dag_k\frac{\partial\hat{\alpha}_k}{\partial t}\hat{r}^{n-1}-\hat{r}^{n-1}\frac{\partial\hat{\alpha}^{\dag}_k}{\partial t}\hat{a}_k\bigg)
			+\sum_{k}\sum_{m=0}^{n-1}\bigg(\bigg[\hat{r}^m,\hat{a}_k^\dag\frac{\partial\hat{\alpha}_k}{\partial t}\bigg]\hat{r}^{n-m-1}-\hat{r}^m\bigg[\frac{\partial \hat{\alpha}_k^{\dag}}{\partial t}\hat{a}_k,\hat{r}^{n-m-1}\bigg]\bigg)
		\end{align}
		and enables us to partly carry out the remaining sum over $n$ in Eq.~\eqref{eq:dispderiv}, yielding
		\begin{equation}
			\frac{\partial \hat{D}}{\partial t}=\sum_{k}\bigg(\hat{a}^\dag_k\frac{\partial\hat{\alpha}_k}{\partial t}\hat{D}-\hat{D}\frac{\partial\hat{\alpha}^{\dag}_k}{\partial t}\hat{a}_k\bigg)
			+\sum_{k}\sum_{n=0}^\infty\frac{1}{n!}\sum_{m=0}^{n-1}\bigg(\bigg[\hat{r}^m,\hat{a}_k^\dag\frac{\partial\hat{\alpha}_k}{\partial t}\bigg]\hat{r}^{n-m-1}-\hat{r}^{n-m-1}\bigg[\frac{\partial \hat{\alpha}_k^{\dag}}{\partial t}\hat{a}_k,\hat{r}^{m}\bigg]\bigg), \label{DdotD}
		\end{equation}
		where we have changed the summation in the very last term according to $m\leftrightarrow n-m-1$. 
		
		We are interested in a theory that is valid up to second order. Therefore, we will systematically neglect higher-order terms. It is useful to write down general expressions for the displaced BM using the Baker-Campbell-Hausdorff relation
		\begin{equation}
			\hat{D}^{\dag}\hat{a}_k\hat{D}=\sum_{l=0}^{\infty}\frac{1}{l!}\left[\sum_{k'}(\hat{\alpha}^{\dag}_{k'}\hat{a}_{k'}-\hat{a}_{k'}^{\dag}\hat{\alpha}_{k'}),\hat{a}_k\right]_l
			\approx\hat{a}_k+\hat{\alpha}_k+\frac{1}{2}\sum_{k'}([\hat{\alpha}_{k'}^{\dag},\hat{\alpha}_{k}]\hat{a}_{k'}-\hat{a}_{k'}^{\dag}[\hat{\alpha}_{k'},\hat{\alpha}_k]),\label{Displaceda}
		\end{equation}
		where $[\cdot,\cdot]_l$ denotes the $l$th nested commutator. In the same manner, for an arbitrary system operator $\hat{O}$ one finds
		\begin{align}
			\hat{D}^{\dag}\hat{O}\hat{D}
			% 	=&\sum_{l=0}^{\infty}\frac{1}{l!}\bigg[\sum_{k'}(\hat{\alpha}^{\dag}_{k'}\hat{a}_{k'}-\hat{a}_{k'}^{\dag}\hat{\alpha}_{k'}),\hat{O}\bigg]_l\nonumber\\
			\approx&\hat{O}+\sum_k([\hat{\alpha}_k^{\dag},\hat{O}]\hat{a}_k+\hat{a}_k^{\dag}[\hat{O},\hat{\alpha}_k])+\frac{1}{2}\sum_k([\hat{\alpha}_k^{\dag},\hat{O}]\hat{\alpha}_k+\hat{\alpha}_k^{\dag}[\hat{O},\hat{\alpha}_k])\nonumber\\
			&+\frac{1}{2}\sum_{k,k'}([\hat{\alpha}^{\dag}_{k'}\hat{a}_{k'}-\hat{a}_{k'}^{\dag}\hat{\alpha}_{k'},[\hat{\alpha}_k^{\dag},\hat{O}]]\hat{a}_k+\hat{a}_k^{\dag}[\hat{\alpha}^{\dag}_{k'}\hat{a}_{k'}-\hat{a}_{k'}^{\dag}\hat{\alpha}_{k'},[\hat{O},\hat{\alpha}_k]]).
		\end{align}
		Applying $D^\dagger$ from the left to the first term of Eq.~\eqref{DdotD} and using the above second-order approximations yields
		\begin{equation}
			\hat{D}^{\dag}\sum_{k}\bigg(\hat{a}^\dag_k\frac{\partial\hat{\alpha}_k}{\partial t}\hat{D}-\hat{D}\frac{\partial\hat{\alpha}^{\dag}_k}{\partial t}\hat{a}_k\bigg)\approx\sum_{k}\bigg(\hat{a}^\dag_k\frac{\partial\hat{\alpha}_k}{\partial t}-\frac{\partial\hat{\alpha}^{\dag}_k}{\partial t}\hat{a}_k\bigg)
			+\sum_{k}\bigg(\hat{\alpha}^\dag_k\frac{\partial\hat{\alpha}_k}{\partial t}-\hat{a}^{\dag}_k\bigg[\hat{r},\frac{\partial\hat{\alpha}_k}{\partial t}\bigg]\bigg).
		\end{equation}
		We now turn to the second term of Eq.~\eqref{DdotD}. We emphasize that $\hat{r}$ itself is of first order and $\partial\hat{\alpha}/\partial t$ is likewise of first order. Since the term with $m=0$ vanishes, the only remaining term is $m=1$ and therefore $n=2$. For the second term in Eq.~\eqref{DdotD} we thereby find 
		\begin{gather}
			\hat{D}^{\dag}\sum_k\sum_{n=0}^{\infty}\frac{1}{n!}\sum_{m=0}^{n-1}\bigg[\hat{r}^m,\hat{a}_k^\dag\frac{\partial\hat{\alpha}_k}{\partial t}\bigg]\hat{r}^{n-m-1}\approx\frac{1}{2}\sum_k\bigg[\hat{r},\hat{a}_k^\dag\frac{\partial\hat{\alpha}_k}{\partial t}\bigg]=\frac{1}{2}\sum_k\bigg(\hat{a}_k^{\dag}\bigg[\hat{r},\frac{\partial \hat{\alpha}_k}{\partial t}\bigg]-\hat{\alpha}^{\dag}_k\frac{\partial \hat{\alpha}_k}{\partial t}\bigg),\\
			\hat{D}^{\dag}\sum_k\sum_{n=0}^{\infty}\frac{1}{n!}\sum_{m=0}^{n-1}\hat{r}^{n-m-1}\bigg[\frac{\partial \hat{\alpha}_k^{\dag}}{\partial t}\hat{a}_k,\hat{r}^{m}\bigg]\approx\frac{1}{2}\sum_k\bigg[\frac{\partial \hat{\alpha}_k^{\dag}}{\partial t}\hat{a}_k,\hat{r}\bigg]=\frac{1}{2}\sum_k\bigg(\bigg[\frac{\partial \hat{\alpha}_k^{\dag}}{\partial t},\hat{r}\bigg]\hat{a}_k+\frac{\partial \hat{\alpha}_k^{\dag}}{\partial t}\hat{\alpha}_k\bigg).
		\end{gather}
		By combining all these terms we finally find
		\begin{equation}
			\mathcal{L}_a\tilde{\rho}=-i\bigg[-i\hat{D}^{\dag}\frac{\partial \hat{D}}{\partial t},\tilde{\rho}\bigg]
		\end{equation}
		with
		\begin{equation}
			\hat{D}^{\dag}\frac{\partial \hat{D}}{\partial t}=\sum_{k}\bigg(\hat{a}^\dag_k\frac{\partial\hat{\alpha}_k}{\partial t}-\frac{\partial\hat{\alpha}^{\dag}_k}{\partial t}\hat{a}_k\bigg)
			+\frac{1}{2}\sum_{k}\bigg(\hat{\alpha}^\dag_k\frac{\partial\hat{\alpha}_k}{\partial t}-\frac{\partial\hat{\alpha}^\dag_k}{\partial t}\hat{\alpha}_k-\bigg[\frac{\partial \hat{\alpha}_k^{\dag}}{\partial t},\hat{r}\bigg]\hat{a}_k-\hat{a}^{\dag}_k\bigg[\hat{r},\frac{\partial\hat{\alpha}_k}{\partial t}\bigg]\bigg). \label{DdotD2}
		\end{equation}

		\subsection{Calculation of $\mathcal{L}_b$}\label{App:A:2}
		We now turn our attention to $\mathcal{L}_b$. This term, as visible in Eq.~\eqref{Displacedmaster}, has a Hamiltonian and a dissipative part. We start with the displaced Hamiltonian that we rewrite as
		\begin{align}
			\hat{D}^{\dag}\hat{H}\hat{D}\approx\tilde{A}+\tilde{B}+\tilde{C}.
		\end{align}
		In the following, we give the explicit forms of $\tilde{A}$, $\tilde{B}$, and $\tilde{C}$ up to second order in the operators $\hat{\alpha}_k$ and $\hat{S}_k$.
		The first term is the displaced QS Hamiltonian $H_S$ and takes the form
		\begin{align}
			\tilde{A}=\hat{D}^{\dag}\hat{H}_S\hat{D}\approx&\hat{H}_S+\sum_k([\hat{\alpha}_k^{\dag},\hat{H}_S]\hat{a}_k+\hat{a}_k^{\dag}[\hat{H}_S,\hat{\alpha}_k])+\frac{1}{2}\sum_k([\hat{\alpha}_k^{\dag},\hat{H}_S]\hat{\alpha}_k+\hat{\alpha}_k^{\dag}[\hat{H}_S,\hat{\alpha}_k])\nonumber\\
			&+\frac{1}{2}\sum_{k,k'}([\hat{\alpha}^{\dag}_{k'}\hat{a}_{k'}-\hat{a}_{k'}^{\dag}\hat{\alpha}_{k'},[\hat{\alpha}_k^{\dag},\hat{H}_S]]\hat{a}_k+\hat{a}_k^{\dag}[\hat{\alpha}^{\dag}_{k'}\hat{a}_{k'}-\hat{a}_{k'}^{\dag}\hat{\alpha}_{k'},[\hat{H}_S,\hat{\alpha}_k]]).
		\end{align}
		The next term is given by the displaced BM Hamiltonian
		\begin{align}
			\tilde{B}=\sum_{k,k'}\hat{D}^{\dag}\hat{a}^{\dag}_{k'}\hat{\Omega}_S^{k',k}\hat{a}_k\hat{D}
			\approx&\sum_{k,k'}\big(\hat{a}^{\dag}_{k'}\hat{\Omega}_S^{k',k}\hat{a}_k+\hat{\alpha}_{k'}^{\dag}\hat{\Omega}_S^{k',k}\hat{a}_k+\hat{a}_{k'}^{\dag}\hat{\Omega}_S^{k',k}\hat{\alpha}_k+\hat{\alpha}^{\dag}_{k'}\hat{\Omega}_S^{k',k}\hat{\alpha}_k\big)\nonumber\\
			&+\sum_{k,k',q}\hat{a}^{\dag}_{k'}\big(\big[\hat{\alpha}_q^{\dag},\hat{\Omega}_S^{k',k}\big]\hat{a}_q+\hat{a}_q^{\dag}\big[\hat{\Omega}_S^{k',k},\hat{\alpha}_q\big]\big)\hat{a}_k\nonumber+\frac{1}{2}\sum_{k,k',q}\hat{a}^{\dag}_{k'}\big(\big[\hat{\alpha}_q^{\dag},\hat{\Omega}_S^{k',k}\big]\hat{\alpha}_q+\hat{\alpha}_q^{\dag}\big[\hat{\Omega}_S^{k',k},\hat{\alpha}_q\big]\big)\hat{a}_k\nonumber\\
			&+\sum_{k,k',q}\hat{\alpha}^{\dag}_{k'}\big(\big[\hat{\alpha}_q^{\dag},\hat{\Omega}_S^{k',k}\big]\hat{a}_q+\hat{a}_q^{\dag}\big[\hat{\Omega}_S^{k',k},\hat{\alpha}_q\big]\big)\hat{a}_k+\frac{1}{2}\sum_{k,k',q}(\hat{a}_{q}^{\dag}[\hat{\alpha}_{k'}^{\dag},\hat{\alpha}_{q}]-[\hat{\alpha}_{k'}^{\dag},\hat{\alpha}_{q}^{\dag}]\hat{a}_{q})\hat{\Omega}_S^{k',k}\hat{a}_k\nonumber\\
			&+\sum_{k,k',q}\hat{a}^{\dag}_{k'}\big(\big[\hat{\alpha}_q^{\dag},\hat{\Omega}_S^{k',k}\big]\hat{a}_q+\hat{a}_q^{\dag}\big[\hat{\Omega}_S^{k',k},\hat{\alpha}_q\big]\big)\hat{\alpha}_k+\frac{1}{2}\sum_{k,k',q}\hat{a}^{\dag}_{k'}\hat{\Omega}_S^{k',k}([\hat{\alpha}_{q}^{\dag},\hat{\alpha}_{k}]\hat{a}_{q}-\hat{a}_{q}^{\dag}[\hat{\alpha}_{q},\hat{\alpha}_k])\nonumber\\
			&+\frac{1}{2}\sum_{k,k',q,q'}\hat{a}^{\dag}_{k'}\big(\big[\hat{\alpha}^{\dag}_{q'}\hat{a}_{q'}-\hat{a}_{q'}^{\dag}\hat{\alpha}_{q'},\big[\hat{\alpha}_q^{\dag},\hat{\Omega}_S^{k',k}\big]\big]\hat{a}_q+\hat{a}_q^{\dag}\big[\hat{\alpha}^{\dag}_{q'}\hat{a}_{q'}-\hat{a}_{q'}^{\dag}\hat{\alpha}_{q'},\big[\hat{\Omega}_S^{k',k},\hat{\alpha}_q\big]\big]\big)\hat{a}_k.
		\end{align}
		We will now explicitly use the fact that the term proportional to $\hat{S}_k$ is already of higher order. Consequently, for the third term, representing the displaced driving, we find 
		\begin{align}
			\tilde{C}=\sum_{k}\hat{D}^{\dag}[\hat{a}^{\dag}_k\hat{S}_k+\hat{S}_k^{\dag}\hat{a}_k]\hat{D}\approx&\sum_{k}(\hat{a}^{\dag}_k\hat{S}_k+\hat{S}_k^{\dag}\hat{a}_k+\hat{\alpha}^{\dag}_k\hat{S}_k+\hat{S}_k^{\dag}\hat{\alpha}_k)\nonumber\\
			&+\sum_{k,k'}\hat{a}^{\dag}_k([\hat{\alpha}_{k'}^{\dag},\hat{S}_k]\hat{a}_{k'}+\hat{a}_{k'}^{\dag}[\hat{S}_k,\hat{\alpha}_{k'}])+\sum_{k,k'}([\hat{\alpha}_{k'}^{\dag},\hat{S}_k^{\dag}]\hat{a}_{k'}+\hat{a}_{k'}^{\dag}[\hat{S}_k^{\dag},\hat{\alpha}_{k'}])\hat{a}_k.
		\end{align}
		Lastly, the form of the displaced master equation in Eq.~\eqref{Displacedmaster} also requires the explicit expressions for the displaced dissipator, which takes the form
		\begin{align}
			\tilde{\mathcal{D}}[\hat{a}_k]\tilde{\rho}=&	\hat{D}^{\dag}(2\hat{a}_k\hat{\rho}\hat{a}^{\dag}_k-\hat{a}^{\dag}_k\hat{a}_k\hat{\rho}-\hat{\rho}\hat{a}^{\dag}_k\hat{a}_k)\hat{D}\nonumber\\
			\approx&(2\hat{a}_k\tilde{\rho}\hat{a}^{\dag}_k-\hat{a}^{\dag}_k\hat{a}_k\tilde{\rho}-\tilde{\rho}\hat{a}^{\dag}_k\hat{a}_k)+(2\hat{a}_k\tilde{\rho}\hat{\alpha}^{\dag}_k-\hat{\alpha}^{\dag}_k\hat{a}_k\tilde{\rho}-\tilde{\rho}\hat{\alpha}^{\dag}_k\hat{a}_k)+(2\hat{\alpha}_k\tilde{\rho}\hat{a}^{\dag}_k-\hat{a}^{\dag}_k\hat{\alpha}_k\tilde{\rho}-\tilde{\rho}\hat{a}^{\dag}_k\hat{\alpha}_k)\nonumber\\
			&+(2\hat{\alpha}_k\tilde{\rho}\hat{\alpha}^{\dag}_k-\hat{\alpha}^{\dag}_k\hat{\alpha}_k\tilde{\rho}-\tilde{\rho}\hat{\alpha}^{\dag}_k\hat{\alpha}_k)+\sum_{k'}\{([\hat{\alpha}_{k'}^{\dag},\hat{\alpha}_{k}]\hat{a}_{k'}-\hat{a}_{k'}^{\dag}[\hat{\alpha}_{k'},\hat{\alpha}_k])\tilde{\rho}\hat{a}^{\dag}_k+\text{H.c.}\}\nonumber\\
			&-\frac{1}{2}\sum_{k'} \{\hat{a}^{\dag}_k([\hat{\alpha}_{k'}^{\dag},\hat{\alpha}_{k}]\hat{a}_{k'}-\hat{a}_{k'}^{\dag}[\hat{\alpha}_{k'},\hat{\alpha}_k])\tilde{\rho}+\text{H.c.}\}-\frac{1}{2}\sum_{k'}\{\tilde{\rho}\hat{a}^{\dag}_k([\hat{\alpha}_{k'}^{\dag},\hat{\alpha}_{k}]\hat{a}_{k'}-\hat{a}_{k'}^{\dag}[\hat{\alpha}_{k'},\hat{\alpha}_k])+\text{H.c.}\}.
		\end{align}
		This concludes the derivation of the displaced master equation up to second order in the coupling and $\hat{\alpha}_k$.
	\end{widetext}
	
	\section{Derivation of the effective Master equation}\label{App:B}
	In this section, we report details on the derivation of the effective master equation, shown in Eq.~(5) of the main text. 
	
	\subsection{Elimination of the bosonic modes}
	We first introduce the projection $\mathcal{P}$ onto the BM vacuum $|\text{vac}\rangle$, with $\mathcal{P}\tilde{\rho}=|\text{vac}\rangle\langle \text{vac}|\tilde{\rho}|\text{vac}\rangle\langle\text{vac}|$, and define the projector onto the orthogonal subspace, given by $\mathcal{Q}=\mathcal{I}-\mathcal{P}$, where $\mathcal{I}$ is the identity superoperator. We now write down the coupled differential equations for the time evolution of $\hat{v}=\mathcal{P}\tilde{\rho}$ and  $\hat{w}=\mathcal{Q}\tilde{\rho}$ that can be written
	\begin{gather}
		\frac{\partial \hat{v}}{\partial t}=\mathcal{P}\left(\mathcal{L}_0+\mathcal{L}_1+\mathcal{L}_2\right)\hat{v}+\mathcal{P}\left(\mathcal{L}_0+\mathcal{L}_1+\mathcal{L}_2\right)\hat{w},\label{dynv}\\
		\frac{\partial \hat{w}}{\partial t}=\mathcal{Q}\left(\mathcal{L}_0+\mathcal{L}_1+\mathcal{L}_2\right)\hat{w}+\mathcal{Q}\left(\mathcal{L}_1+\mathcal{L}_2\right)\hat{v},\label{dynw}
	\end{gather}
	where the superoperators $\mathcal{L}_{i}$ include driving due to $\hat{S}_k$ and displacements $\hat{\alpha}_k$ ($\hat{\alpha}_k^{\dag}$) up to the $i$th order. Due to this hierarchy and the results of the previous section, we find the coupling between $\hat{v}$ and $\hat{w}$ to be at least of first order, resulting in $\mathcal{Q}\mathcal{L}_0\hat{v}=0$. However, there is a first-order contribution coupling the dynamics of $\hat{v}$ and $\hat{w}$ that is mediated by the BM raising operators $\hat{a}_k^{\dag}$. This term is given by
	\begin{equation}
		\mathcal{Q}\mathcal{L}_1\hat{v}=\sum_k\hat{E}_k|1_k\rangle\langle\text{vac}|\hat{\rho}_{\text{sys}}+\text{H.c.},
	\end{equation}
	in which $|1_k\rangle=\hat{a}^\dagger_k|\text{vac}\rangle$ represents the state where the $k$th BM carries one excitation and all remaining BM are in the vacuum. Here, we have introduced $\hat{\rho}_{\text{sys}}=\langle\text{vac}|\tilde{\rho}|\text{vac}\rangle$ and the explicit form of $\hat{E}_k$ reads
	\begin{equation}
		\label{eq:ck}
		\hat{E}_k=-\frac{\partial \hat{\alpha}_k}{\partial t}-i[\hat{H}_S,\hat{\alpha}_k]-i\sum_{k'}\hat{\Omega}_S^{k,k'}\hat{\alpha}_{k'}-i\hat{S}_k-\kappa_k\hat{\alpha}_k.
	\end{equation}
	Since we have not yet specified the actual form of the $\hat{\alpha}_k$, we can simply impose the condition
	\begin{align}
		\hat{E}_k=0. \label{alphak}
	\end{align}
	for every $k$. We remark that in the no-driving case, $\hat{S}_k=0$, one solution of this equation is simply $\hat{\alpha}_k=0$, which is, therefore, consistent with the fact that the BM degrees of freedom relax to the vacuum state. Enforcing these conditions in Eq.~\eqref{alphak} results in
	\begin{equation}
		\mathcal{Q}\mathcal{L}_1\hat{v}=0.
	\end{equation}
	This means the contribution of $\hat{v}$ to $\hat{w}$ is at least of second order. Consequently, we find that the only contribution of $\hat{w}$ to $\hat{v}$ is due to the coupling
	\begin{equation}
		\mathcal{P}\mathcal{L}_0\hat{w}=2\sum_k\kappa_k|\text{vac}\rangle\langle 1_k|\hat{w}|1_k\rangle\langle \text{vac}|.
	\end{equation}
	However, there is no second order term that couples the vacuum state to $|1_k\rangle\langle 1_k|$ in the dynamics of $\hat{w}$. Therefore, this term is at least of third order and thus negligible. 
	
	Hence, we have shown that with our specific choice of $\hat{\alpha}_k$, determined by Eqs.~\eqref{eq:ck} and~\eqref{alphak}, we have decoupled $\hat{v}$ and $\hat{w}$ up to third-order corrections. 
	The master equation can thereby be written as
	\begin{equation}
		\frac{\partial \hat{v}}{\partial t}=\mathcal{P}\mathcal{L}_0\hat{v}+\mathcal{P}\mathcal{L}_1\hat{v}+\mathcal{P}\mathcal{L}_2\hat{v},\label{dynv2}
	\end{equation}
	and the BM are effectively eliminated. The effective QS master equation can now be obtained by tracing out the BM degrees of freedom.
	
	\subsection{Effective master equation for the atoms}
	Applying the partial trace over the BM degrees of freedom, $\text{Tr}_{\text{BM}}$, onto Eq.~\eqref{dynv2}, we find that the effective master equation for $\hat{\rho}_{\text{sys}}$ is given by
	\begin{equation}
		\frac{\partial\hat{\rho}_{\text{sys}}}{\partial t}=-i[\hat{H}_{\text{eff}},\hat{\rho}_{\text{sys}}]+\sum_k\kappa_k\mathcal{D}[\hat{\alpha}_k]\hat{\rho}_{\text{sys}}\label{effMaster}
	\end{equation}
	with the effective Hamiltonian
	\begin{align}
		\hat{H}_{\text{eff}}=\hat{H}_S&-\frac{i}{2}\sum_{k}\left(\hat{\alpha}^\dag_k\frac{\partial\hat{\alpha}_k}{\partial t}-\frac{\partial\hat{\alpha}^\dag_k}{\partial t}\hat{\alpha}_k\right)\nonumber\\
		&+\sum_k\Bigg(\frac{[\hat{\alpha}_k^{\dag},\hat{H}_S]\hat{\alpha}_k+\hat{\alpha}_k^{\dag}[\hat{H}_S,\hat{\alpha}_k]}{2}\nonumber\\
		&+\sum_{k'}\hat{\alpha}^{\dag}_{k}\hat{\Omega}_S^{k,k'}\hat{\alpha}_{k'}+\hat{\alpha}^{\dag}_k\hat{S}_k+\hat{S}_k^{\dag}\hat{\alpha}_k\Bigg).
	\end{align}
	Now, using the elimination condition~\eqref{alphak} in the form 
	\begin{equation}
		\label{alphak2}
		\frac{\partial \hat{\alpha}_k}{\partial t}	=-i[\hat{H}_S,\hat{\alpha}_k]-i\sum_{k'}\hat{\Omega}_S^{k,k'}\hat{\alpha}_{k'}-i\hat{S}_k-\kappa_k\hat{\alpha}_k
	\end{equation}
	as well as its Hermitian-conjugate version, the effective Hamiltonian above can be rewritten as
	\begin{equation}
		\hat{H}_{\text{eff}}=\hat{H}_S+\frac{1}{2}\sum_{k}(\hat{\alpha}^{\dag}_k\hat{S}_k+\hat{S}_k^{\dag}\hat{\alpha}_k).\label{Heff2}
	\end{equation}
	
	\section{Atom-only mean-field description}\label{App:C}
	In this section, we derive the atom-only mean-field description of the dynamics for the dissipative Dicke model. As mentioned in the main text, we use we use $\hat{H}_S=\omega_0 \hat{S}^{z}$
	and describe the coupling by $\hat{\Omega}_S=\omega_c$ and $\hat{S}=2g\hat{S}^x/\sqrt{N}$. We then rewrite the explicit form of $\alpha_{\pm}$ in an alternative form given by
	\begin{equation}
		\hat{\alpha}\approx-\frac{2g\hat{S}^x}{\sqrt{N}(\omega_c-i\kappa)}+\frac{2ig\omega_0\hat{S}^y}{\sqrt{N}(\omega_c-i\kappa)^2}.
	\end{equation}
	The effective Hamiltonian, as given in Eq.~\eqref{Heff2}, then reads
	\begin{align}
		\hat{H}_{\text{eff}}=&\omega_0 \hat{S}^{z}-\frac{4\omega_c g^2}{N(\omega_c^2+\kappa^2)}[\hat{S}^x]^2-\frac{4\omega_c \kappa \omega_0g^2}{N(\omega_c^2+\kappa^2)^2} \{\hat{S}^x,\hat{S}^y\}\nonumber\\
		&-\frac{2g^2\omega_0(\omega_c^2-\kappa^2)}{N(\omega_c^2+\kappa^2)^2}\hat{S}^z.\label{HeffDicke}
	\end{align}
	For the remainder of this section, we will focus on the limit where $N\to\infty$. In this case, the three terms in the first line of Eq.~\eqref{HeffDicke} scale as $N$ while the term in the second line scales as $1$ and is therefore negligible. Using the effective Hamiltonian and the effective dissipator in this limit results in the equations
	\begin{gather}
		\frac{d \hat{S}^x }{dt}=i[\hat{H}_\text{eff},\hat{S}^x]+\kappa(\hat{\alpha}^{\dag}[\hat{S}^x,\hat{\alpha}]+[\hat{\alpha}^{\dag},\hat{S}^x]\hat{\alpha})
		\approx-\omega_0\hat{S}^y,\\
		\frac{d\hat{S}^y}{dt}
		\approx\omega_0\hat{S}^x+\frac{2V_0}{N}\{\hat{S}^x,\hat{S}^z\}+\frac{2V_1}{N}\{\hat{S}^y,\hat{S}^z\},\\
		\frac{d\hat{S}^z}{dt}
		\approx-\frac{2V_0}{N}\{\hat{S}^x,\hat{S}^y\}-\frac{4V_1}{N}\hat{S}^y\hat{S}^y,
	\end{gather}
	with $V_0=2\omega_cg^2/(\omega_c^2+\kappa^2)$ and $V_1=4\omega_c\kappa\omega_0 g^2/(\omega_c^2+\kappa^2)^2$.
	We now take the expectation value $\langle\,\cdot\,\rangle$ and make a mean-field approximation where we factorize second moments $\langle\hat{S}^a\hat{S}^b\rangle=\langle\hat{S}^a\rangle\langle\hat{S}^b\rangle$. Then, we replace the operators by their $c$-number equivalents $S^a=\langle\hat{S}^a\rangle$, which obey the coupled non-linear differential equations given by
	\begin{gather}
		\frac{dS^x}{dt}=-\omega_0S^y,\nonumber\\
		\frac{dS^y}{dt}=\omega_0S^x+\frac{4V_0}{N}S^xS^z+\frac{4V_1}{N}S^yS^z,\nonumber\\
		\frac{dS^z}{dt}=-\frac{4V_0}{N}S^xS^y-\frac{4V_1}{N}S^yS^y.\label{SxSySz}
	\end{gather}
	These equations are the same as the ones reported in Eq.~(35) of Ref.~\cite{Damanet:2019} and, therefore, give rise to the same steady state and the same oscillation as well as damping rates.
	
	\section{Dynamical critical exponent}
	In this section, we show that the mean-field equations provide the correct critical exponents. For this, we assume that the system is in a stable stationary state with $S^y=0$, $S^x=Nx/2$, and $S^z=-Nz/2$, where $x^2+z^2=1$. Below threshold, $g<g_c$, we have $z=1$, whereas above threshold, $g>g_c$, we have $z=\omega_0/2V_0$ (see, e.g., Ref.~\cite{Damanet:2019}). We now study the dynamics of fluctuations $\delta S^x$, $\delta S^y$, and $\delta S^z$ to determine the stability and relaxation of this configuration. Defining $\boldsymbol{\delta}=(\delta S^x, \delta S^y, \delta S^z)^T$ we obtain
	\begin{align}
		\frac{d\boldsymbol{\delta}}{dt}={\bf M}\boldsymbol{\delta}
	\end{align}
	with
	\begin{align}
		{\bf M}=\begin{pmatrix}
			0&-\omega_0&0\\
			\omega_0-2V_0z&-2V_1z&+2V_0x\\
			0&-2V_0x&0
		\end{pmatrix}.
	\end{align}
	Below threshold, $g<g_c$ and $x=0$, the eigenvalues are the solutions of
	\begin{align}
		\lambda(\lambda+2V_1)+\omega_0(\omega_0-2V_0)=0, 
	\end{align}
	which are given by
	\begin{align}
		\lambda=-V_1\pm\sqrt{V_1^2-\omega_0(\omega_0-2V_0)}.
	\end{align}
	Since $\omega_0-2V_0\propto g_c-g$ this gives the critical exponent $\nu=1$ with $\lambda\propto|g_c-g|^\nu$ for non-vanishing $V_1$. On the other hand, if $V_1=0$ we obtain the critical exponent $\nu = 0.5$ that is known from the Dicke quantum phase transition.
	
	Above threshold, $g>g_c$, we need to solve
	\begin{align}
		\lambda(\lambda+2V_1z)+4V_0^2x^2=0
	\end{align}
	and the solutions are given by
	\begin{align}
		\lambda=-V_1z\pm\sqrt{V_1^2z^2-4V_0^2x^2}.
	\end{align}
	Since $z\approx1$ at the critical point and $x=\pm(g^4-g_c^4)^{1/2}/g^2\propto|g-g_c|^{1/2}$, we obtain $\lambda\propto|g_c-g|^\nu$ with a critical exponent $\nu=1$ for non-vanishing $V_1$ and an exponent $\nu = 0.5$ for $V_1=0$.
	
	This calculation shows that the mean-field equations derived from the effective Lindblad master eqauation give the correct dynamical critical exponent $\nu=1$ of the dissipative Dicke model~\cite{Nagy:2011,DallaTorre:2013,Brennecke:2013} and also includes the correct exponent $\nu=0.5$ in absence of dissipation, i.e., $V_1=0$.
	
	\section{Stochastic Simulation of the dissipative Dicke model}\label{App:D}
	The purpose of this section is to give the explicit form of the stochastic differential equation that we compare with the simulation of the effective master equation.  We first start by writing down the Heisenberg-Langevin equations that are given by
	\begin{gather}
		\frac{d\hat{a}}{dt}=-\kappa\hat{a}-i\omega_c\hat{a}-i2\frac{g}{\sqrt{N}}\hat{S}^x+\sqrt{2\kappa}\hat{a}_{\text{in}}(t),\\
		\frac{d\hat{S}^x}{dt}=-\omega_0\hat{S}^y,\\
		\frac{d\hat{S}^y}{dt}=\omega_0\hat{S}^x-2\frac{g}{\sqrt{N}}(\hat{a}+\hat{a}^{\dag})\hat{S}^z,\\
		\frac{d\hat{S}^z}{dt}=2\frac{g}{\sqrt{N}}(\hat{a}+\hat{a}^{\dag})\hat{S}^y.
	\end{gather}
	Here, we have introduced the noise operators $\hat{a}_{\text{in}}$ with $\langle\hat{a}_{\text{in}}\rangle=0$ and the second moments $\langle\hat{a}_{\text{in}}(t)\hat{a}_{\text{in}}(t')\rangle=0=\langle\hat{a}^{\dag}_{\text{in}}(t)\hat{a}_{\text{in}}(t')\rangle$ as well as $\langle\hat{a}_{\text{in}}(t)\hat{a}^{\dag}_{\text{in}}(t')\rangle=\delta(t-t')$. The average is here taken over the modes external to the cavity mode. 
	
	We now define the two quadrature operators $\hat{x}=\hat{a}^{\dag}+\hat{a}$ and $\hat{p}=i(\hat{a}^{\dag}-\hat{a})$ and write down the equations of motion, which have the form
	\begin{gather}
		\frac{d\hat{x}}{dt}=-\kappa\hat{x}+\omega_c\hat{p}+\sqrt{2\kappa}\hat{\mathcal{N}}^x(t),\\
		\frac{d\hat{p}}{dt}=-\kappa\hat{p}-\omega_c\hat{x}-4\frac{g}{\sqrt{N}}\hat{S}^x+\sqrt{2\kappa}\hat{\mathcal{N}}^p(t),\\
		\frac{d\hat{S}^x}{dt}=-\omega_0\hat{S}^y,\\
		\frac{d\hat{S}^y}{dt}=\omega_0\hat{S}^x-2\frac{g(t)}{\sqrt{N}}\hat{x}\hat{S}^z,\\
		\frac{d\hat{S}^z}{dt}=2\frac{g}{\sqrt{N}}\hat{x}\hat{S}^y,
	\end{gather}
	with $\hat{\mathcal{N}}^x=(\hat{a}_{\text{in}}(t)+\hat{a}_{\text{in}}^{\dag}(t))$ and $\hat{\mathcal{N}}^p=-i(\hat{a}_{\text{in}}(t)-\hat{a}_{\text{in}}^{\dag}(t))$. The stochastic differential equations are now derived by replacing the quantum operators with $c$-numbers with symmetric ordering and the quantum noise with classical noise having the correct second moments~\cite{Domokos:2001}. The result reads
	\begin{gather}
		\frac{dx}{dt}=-\kappa x+\omega_c p+\sqrt{2\kappa}{\mathcal{N}}^x(t),\\
		\frac{dp}{dt}=-\kappa p-\omega_cx-4\frac{g}{\sqrt{N}}{S}^x+\sqrt{2\kappa}{\mathcal{N}}^p(t),\\
		\frac{dS^x}{dt}=-\omega_0{S}^y,\\
		\frac{dS^y}{dt}=\omega_0{S}^x-2\frac{g}{\sqrt{N}}x{S}^z,\\
		\frac{dS^z}{dt}=2\frac{g}{\sqrt{N}}x{S}^y,
	\end{gather}
	with the classical noise $\mathcal{N}^a$ fulfilling $\langle \mathcal{N}^a\rangle=0$ and $\langle \mathcal{N}^a(t)\mathcal{N}^b(t')\rangle=\delta(t-t')$.
	
	\begin{figure}[tb]
		\center
		\includegraphics[width=0.9\linewidth]{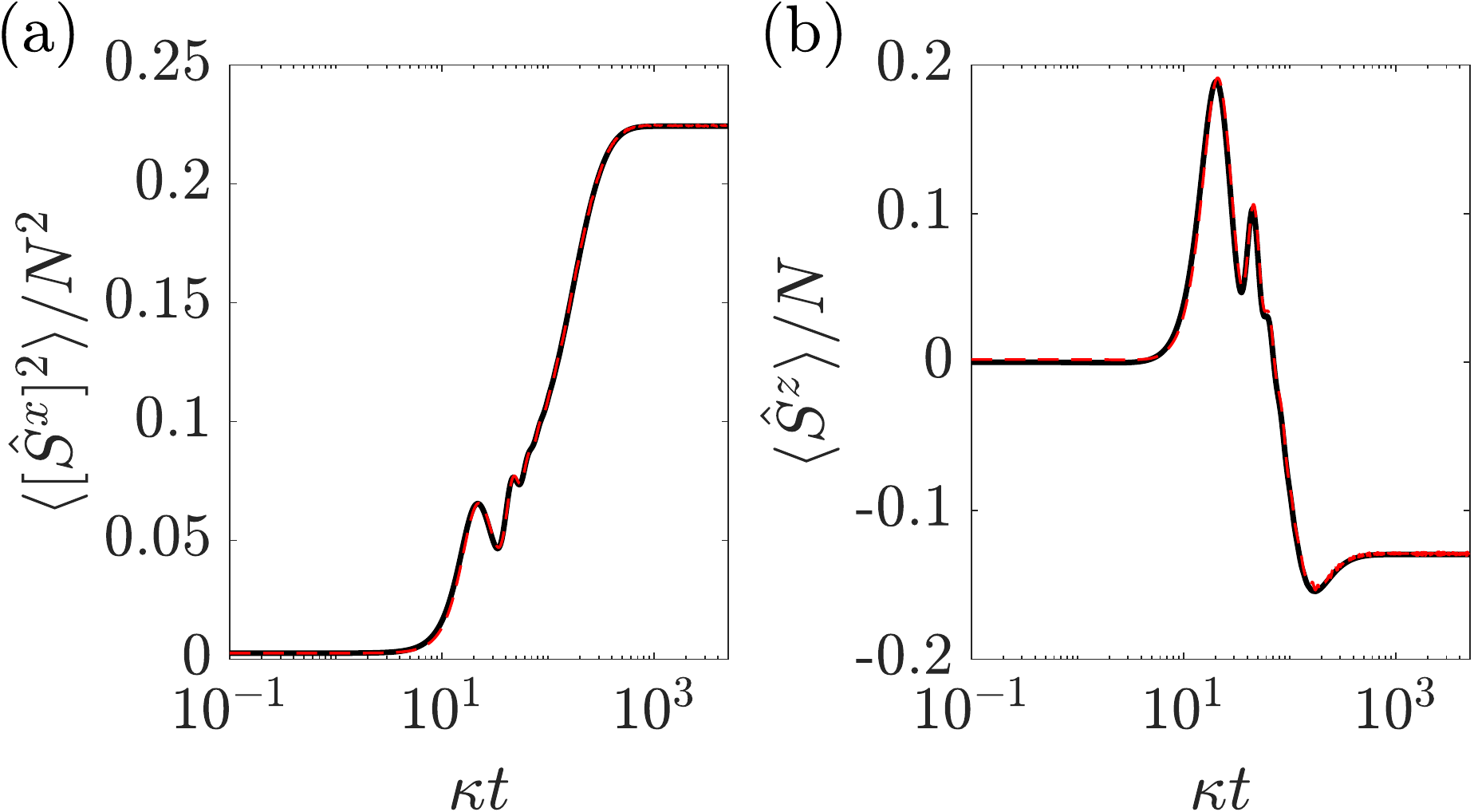}
		\caption{Dynamics of $\langle[\hat{S}^x]^2\rangle$ (a) and  $\langle\hat{S}^z\rangle$ (b) for $g=2g_c$ when the atoms are initialized in the NOON state. The black lines are obtained by simulating the effective master equation~\eqref{effMaster} with $N=100$. The red dashed lines are simulated with the stochastic method, averaged over 20000 simulations with $N=100$. The remaining parameters are $\omega_c=\kappa$, $\omega_0=0.1\kappa$.\label{FigS:1}}
	\end{figure}
	
	\subsection{Z-polarized State}
	In the Letter, we show the dynamics of the dissipative Dicke model with stochastic differential equations when all atoms are initialized in the atomic ground state. This state is $|\Psi\rangle=|-N/2\rangle$. For this, in the thermodynamic limit $N\to\infty$, the correct description is
	$S^z(t=0)=-N/2$ and $S^x(0)$ and $S^y(0)$ are independent random variables that are sampled from a Gaussian with zero mean and variance $N/2$. Notice that this fulfills
	\begin{align}
		\langle[\hat{S}^x]^2+[\hat{S}^y]^2+[\hat{S}^z]^2\rangle=\frac{N}{2}\left(\frac{N}{2}+1\right).
	\end{align}
	The field is initialized in its vacuum state. In the stochastic equation, this is realized by sampling $x$ and $p$ from independent Gaussian distributions with zero mean and unit variance.

	\subsection{NOON State}
	In the Letter, we also study the dynamics of the NOON state $|\Psi\rangle=(|-N/2\rangle+|N/2\rangle)/\sqrt{2}$. In order to sample this state we initialize, as for the polarized state, $S^x(0)$ and $S^y(0)$ as independent random variables with zero mean and variance $N/2$. Since $\langle\hat{S}^z(0)\rangle=0$ but $\langle[\hat{S}^z]^2(0)\rangle=N^2/4$, we also sample $S^z$ from a distribution with zero mean and variance $N^2/4$. This distribution is chosen to have two ``delta'' peaks at $m=-N/2$ and $m=N/2$ states. At this point, we mention that this cannot be distinguished from the mixed state $\hat{\rho}_{\text{sys}}(0)=|N/2\rangle\langle N/2|/2+|-N/2\rangle\langle -N/2|/2$, which does not include the coherences $\hat{c}$ that are mentioned in the Letter. Using this initial condition together with the vacuum state of the cavity, we can now evolve the system. 
	
	The red dashed lines in Fig.~S\ref{FigS:1} show the results of this simulation. In addition, we also plot the simulation of the effective master equation visible as black lines. Both numerical simulations are in excellent agreement. We remark that the stochastic simulation is able to predict the correct moments for the $\hat{S}^a$ spin operators, however, it is not able to predict the fast oscillations visible in Fig.~4 of the Letter.

\end{document}